\documentclass{aa}  

\usepackage{graphicx}
\usepackage{txfonts}
\usepackage{mathrsfs}
\usepackage[colorlinks,allcolors=blue]{hyperref}
\usepackage{breakurl}
\usepackage{subfigure}
\def \inte {INTEGRAL}

\def \sw {{\it Swift}}
\def \chandra {{\it Chandra}}
\def \suz {\emph{Suzaku}}

\def \src {\mbox{RT~Cru}}

\begin{document}

     \title{RT Cru: a look into the X-ray emission of a peculiar symbiotic star}


   \author{L. Ducci
          \inst{1,2}
          \and 
          V. Doroshenko\inst{1}
          \and
          V. Suleimanov\inst{1,3}
          \and
          M. Niko\l ajuk
          \inst{4,2}
          \and
          A. Santangelo
          \inst{1}
          \and
          C. Ferrigno
          \inst{2}
          }

   \institute{Institut f\"ur Astronomie und Astrophysik, Kepler Center for Astro and Particle Physics, Eberhard Karls Universit\"at, 
              Sand 1, 72076 T\"ubingen, Germany\\
              \email{ducci@astro.uni-tuebingen.de}
              \and
              ISDC Data Center for Astrophysics, Universit\'e de Gen\`eve, 16 chemin d'\'Ecogia, 1290 Versoix, Switzerland
              \and
              Kazan Federal University, Kremlevskaja str., 18, Kazan 420008, Russia              
              \and
              Wydzia\l\ Fizyki, Uniwersytet w Bia\l ymstoku, Cio\l kowskiego 1L, 15-245 Bia\l ystok, Poland
   }

   \date{Received ...; accepted ...}

 
  \abstract
   {Symbiotic stars are a heterogeneous class of interacting binaries.
    Among them, \src\ has been classified as prototype of a subclass that is characterised by
    hard X-ray spectra extending past $\sim 20$~keV.
    We analyse $\sim8.6$~Ms of archival \inte\ data collected in the period 2003-2014,
    $\sim140$~ks of \sw/XRT data, and a \suz\ observation of 39\,ks,
    to study the spectral X-ray emission and investigate the nature of the compact object.
Based on the 2MASS photometry, we estimate the distance to the source
of 1.2$-$2.4~kpc. 
The X-ray spectrum obtained with \sw/XRT, JEM-X, IBIS/ISGRI, and \suz\ data
is well fitted by a cooling flow model modified by an absorber that fully covers the source
and two partial covering absorbers. Assuming that the 
hard X-ray emission of \src\ originates from an optically
thin boundary layer around a non-magnetic white dwarf, we estimated 
a mass of the WD of $M_{\rm WD} \approx 1.2$~M$_\odot$.
The mass accretion rate obtained for this source might be too high for 
the optically thin boundary layer scenario. 
Therefore we investigate other plausible scenarios to model its hard X-ray emission.
We show that, alternatively, the observed X-ray spectrum can be explained with the 
X-ray emission from the post-shock region above the polar caps 
of a magnetised white dwarf with mass $M_{\rm WD} \approx 0.9-1.1$~M$_\odot$.}


   \maketitle
%

\section{Introduction}

Symbiotic stars are interacting binaries discovered at early 1900's
and defined as objects whose optical spectra show bright emission lines
(indicating the presence of a hot source) superimposed on a late-type
stellar continuum (commonly a M giant; e.g. \citealt{Allen84}; \citealt{Luna13}).
This vague definition makes symbiotic stars a very heterogeneous
class of objects.
RT~Crucis (hereafter \src), which is the subject of this paper,
is a member of this class.

\src\ was discovered as a variable star (\citealt{Leavitt06}; \citealt{AstNach1911}) and later
classified as a symbiotic star by \citet{Cieslinski94}.
It shows optical flickering in the V and B bands on timescales
of 10-30 minutes and amplitudes up to 0.09 magnitudes. 
Optical flickering has been observed in other symbiotic stars
with accreting white dwarfs (e.g. CH~Cyg, CI~Cyg, T~CrB, RS~Oph, 
\citealt{Cieslinski94}) and accreting neutron stars (GX~1+4, \citealt{Braga93, Jablonski97}).
\citet{Cieslinski94} noted that the optical spectrum of \src\ shows strong similiraties with that 
of the recurrent nova T~CrB in quiescence. 
A search in historical data revealed a large magnitude variation
($\Delta m \sim 1.7$ in \citealt{Leavitt06}; $\Delta m \sim 2.4$ in \citealt{Uiterdijk60})
compatible with those observed in other symbiotic stars \citep{Cieslinski94}.
\citet{Gromadzki13} detected two periodicities 
in ASAS, OGLE, and MACHO lightcurves, $P_{\rm orb}=325\pm9$~d and $P_{\rm puls}=63\pm1$~d,
interpreted as the orbital period and the pulsation of the red giant.
More than a hundred year after its discovery, \src\ 
has recently acquired a great interest, after the observation of 
hard X-ray emission detected up to $\sim 150$~keV (\citealt{Chernyakova05} ; \citealt{Kennea09}).
\src\ was observed by \inte\ in 2003 and 2004
with a 20$-$60~keV flux of $\approx 4\times 10^{-11}$~erg~cm$^{-2}$~s$^{-1}$ \citep{Chernyakova05}
and initially classified as a new source (IGR~J12349$-$6434).
\citet{Masetti05} noted that the position of IGR~J12349$-$6434 was compatible with 
that of \src. The association between these two sources was later confirmed
by \citet{Tueller05} thanks to the refined position of IGR~J12349$-$6434 obtained
with \sw/XRT.
\src\ was detected again with \inte\ with a 20$-$40~keV flux 
$F_{\rm x} > 4\times 10^{-11}$~erg~cm$^{-2}$~s$^{-1}$ \citep{Sguera12,Sguera15}.
The bright, hard X-ray emission properties of \src\ are shared with
a small group of symbiotic stars: T~CrB, CD$-$57~3057, and CH~Cyg \citep{Luna07}.
In particular, the recurrent nova T~CrB, which is believed
to be a non-magnetic $~1.3$\,M$_\odot$ WD where the hard X-ray emission is produced in the
optically thin boundary layer of the accretion disc, has a high X-ray luminosity similar to \src\
($L_{\rm x}\approx 10^{34}$\,erg\,s$^{-1}$, e.g. \citealt{Kennea09,Luna08}).
\citet{Luna07} reported on a \chandra\ observation of \src.
They proposed that, similarly to T~CrB, 
the $0.3-8$~keV X-ray emission and the flickering variability
originates from an optically thin boundary layer around 
a massive non-magnetic white dwarf (WD) powered by an accretion disc.
\citet{Kennea09} studied \src\
using the data from XRT and BAT instruments on board \sw\
and suggested that the X-ray variability on timescales of few days
is caused by variable absorption from a clumpy medium moving across the line of sight.
\citet{Eze14} studied the iron line complex at $6.4-7$~keV of \src\ with \emph{Suzaku}.
He proposed that the 6.4~keV emission line (Fe\,K$\alpha$) 
is caused by the interaction between the hard X-rays 
and the absorbing material around the system, 
while the 6.7~keV (Fe~XXV) and 7~keV (Fe~XXVI) emission lines
are produced by photoionization and collisional excitations in the hot plasma
in the vicinity of the WD.

In this work we coherently analyse the largest possible set of X-ray data
($\sim12$ years of archival \inte\ data, $\sim140$~ks of \sw/XRT data
and $\sim39$~ks of \suz\ observation) to further investigate the
properties of \src\ (Sect. \ref{Sect. obs}) and clarify the nature
of the compact object (Sect. \ref{sect. discussion}).

\section{Observations and data analysis}
\label{Sect. obs}

\subsection{INTEGRAL}
We analysed archival data of the 
INTErnational Gamma-Ray Astrophysics Laboratory (INTEGRAL, \citealt{Winkler03})
collected with the coded-mask telescope
Imager on board INTEGRAL Satellite 
(IBIS, \citealt{Ubertini03}) and the detector
INTEGRAL Soft Gamma-Ray Imager
(ISGRI, \citealt{Lebrun03}). 
IBIS/ISGRI has a fully-coded field of view of $9^\circ \times 9^\circ$,
a partially coded field of view of $29^\circ \times 29^\circ$.
and operates in the $\sim 15-400$~keV energy band.
We also used data from the X-ray monitors 
Joint European Monitor for X-ray
(JEM-X, \citealt{Lund03}),
which are two co-aligned coded-mask telescopes 
on board \inte\ operating simultaneously with IBIS/ISGRI 
in the energy range $3-35$~keV. 
Their fields of view have a diameter of $13.2^\circ$.
Each orbital revolution of INTEGRAL consists of pointings called Science Windows (ScWs).
We performed the reduction and data analysis using the Off-line Science 
Analysis (OSA) 10.2 software provided by the ISDC Data Centre for Astrophysics
\citep{Goldwurm03, Courvoisier03}.
We analysed the data between 2003 January 29 and 2014 December 20 where \src\ was within
12$^\circ$  from the centre of the IBIS/ISGRI field of view,
because at larger off-axis angle the IBIS response is not well known\footnote{See the \inte\
data analysis documentation \url{http://www.isdc.unige.ch/integral/analysis}}.
Data were filtered to exclude bad time intervals. 
The resulting data set consists of 2880 ScWs, 
corresponding to an exposure time of 8.6~Ms.

\src\ was never detected, being below the $5\sigma$ threshold of detection,
in the individual $22-50$~keV sky images of each pointing,
while it was detected with a significance of $\sim 30\sigma$
in the total mosaic image obtained from the combination of the individual images of the whole data set.

We extracted the average JEM-X1 and IBIS/ISGRI spectra from the mosaics
using the OSA tool {\tt mosaic\_spec}, which is especially suitable for
faint sources. Due to the poor statistic,
we did not extract the average JEM-X2 spectrum (the source was observed only in a few pointings).
We used the exposure-weighted average ancillary response files and the rebinned response matrices
specifically generated for the IBIS/ISGRI and JEM-X datasets.
Before fitting, we added systematic uncertainties of 2\% and 3\% to the IBIS/ISGRI and JEM-X
spectra, respectively.

\subsection{Swift/XRT}

\src\ was observed with the X-ray Telescope (XRT, \citealt{Burrows05})
on board the \sw\ satellite \citep{Gehrels04} since 2005.
We collected all the available XRT observations of the target
in the time period from 2005 August 20 23:43:02 UTC
to 2012 December 24 10:45:58 UTC (38 observations, total exposure time of $\sim$140~ks).
Part of these data (first six observations) were analysed and the results presented in \citet{Kennea09}.
We processed the XRT data obtained in photon-counting (PC) mode
with the standard procedures ({\tt xrtpipeline} v0.13.1)
and the calibration files CALDB 4.5.6.
Standard grade filtering (0-12) and screening criteria were applied.
We checked that no pile-up correction was required.
Source counts were accumulated from a circular region
with a radius of 20 pixels (1 pixel $\sim2.36^{\prime\prime}$) 
and background counts were accumulated from a circular nearby source-free region
at RA(J2000)=12:34:17.84, Dec(J2000)=-64:32:08.7 with a
radius of 40 pixels.
The spectral energy channels are grouped to have at least 30 counts per bin
for good statistical quality of the spectrum.

\subsection{Suzaku}
\label{sect. suzaku}

\suz\ \citep{Mitsuda07} observed \src\ in 2007 July 2 ($\sim 39$~ks).
We used data from the X-ray Imaging Spectrometers (XIS, \citealt{Koyama07})
and the PIN layer of the Hard X-ray Detector (HXD \citealt{Takahashi07}),
which were operated in standard mode during the observation.

The data were processed with the standard \suz\ pipeline
included in the HEASOFT software package (version 6.16)
and screened according to standard criteria.

XIS events of \src\ were extracted in each of the three detectors
(front-illuminated: XIS\,0,\,3; back-illuminated: XIS\,1)
in circular regions with radius 2$^\prime$ centered on the target.
Background counts were accumulated from a source-free circular region
far from the point-spread function of the source
(coordinates of the background region:
RA(J2000)=12:35:01.81, Dec(J2000)=-64:41:56.2, radius=2$^\prime$).
Ancillary response and redistribution matrices files for the XIS
spectra were generated using  the \suz\ tools {\tt xisrmfgen}
and {\tt xissimarfgen}.
For the background subtraction in PIN data we used the
\emph{tuned} Non X-ray Background (NXB) models specifically generated
for these two observations by the HXD team.
For the Cosmic X-ray Background (CXB) we used the simulated
spectrum generated by the \suz\ tool {\tt hxdpinxbpi}.
The PIN response files appropriate for the epoch of the observations
and for a point source were selected.
The XIS and PIN spectra were rebinned using the optimal binning method
of \citet{Kaastra16}.
For the timing analysis we correct the event arrival times 
to the Solar system barycenter 
with the \suz\ specific tool {\tt aebarycen}.

\section{Results}
\label{sect. results}

\subsection{Distance}
\label{sect. distance}

We compared the published optical/near-infrared spectrum
of \src\ (fig. 2 of \citealt{Cieslinski94}) with 
those of the spectral atlas of \citet{Pickles98}.
Based on the general shape of the spectra, we classified
the donor star of \src\ as a M4III$-$M5III, that is in agreement with
the M4$-$M5 spectral type obtained by \citet{Cieslinski94}.
Since the optical part of the spectrum might be affected by the emission
from the compact object companion \citep{Hellier01}, we estimated the distance using
the infrared magnitudes of the 2MASS \citep{Skrutskie06} 
counterpart 2MASS~J12345374$-$6433560.
We converted the observed colours and magnitudes to the photometric system
of \citet{Bessell88} using the transformation equations of 
\citet{Carpenter01}\footnote{In particular, we adopted the updated version
of the transformation equations of \citet{Carpenter01},
that can be found in: \url{http://www.astro.caltech.edu/~jmc/2mass/v3/transformations/}}.
Using these information and the intrinsic colours and magnitudes of M4III$-$M5III
stars from \citet{Pickles98} and \citet{Fluks98}, we calculated the interstellar
reddening $E(J-K)$. We obtained the extinction $A_\lambda$ (where $\lambda = J$, $H$, $K$)
using the formula of \citet{Draine89}, $A_\lambda/E(J-K)=2.4(\lambda/\mu m)^{-\beta}$
(which is valid in the range $0.9<\lambda<6$~$\mu$m, and $\beta$
ranges from $1.70$ to $1.80$, \citealt{Draine03}).
M giants show intrinsic variability.
Therefore, we added a maximum amplitude variation of $\Delta$mag$=0.25$ (\citealt{Eyer08}; \citealt{Tabur09})
to the 2MASS magnitude uncertainties of \src.
We found that the distance of \src\ is in the range $\approx 1.2-2.4$~kpc.
This result is in agreement with the value $1.5-2$~kpc reported
in \citet{Luna07} from a private communication
of J.~Miko\l ajewska (2006).

\subsection{Lightcurve analysis}

\begin{figure}
\begin{center}
\includegraphics[width=\columnwidth]{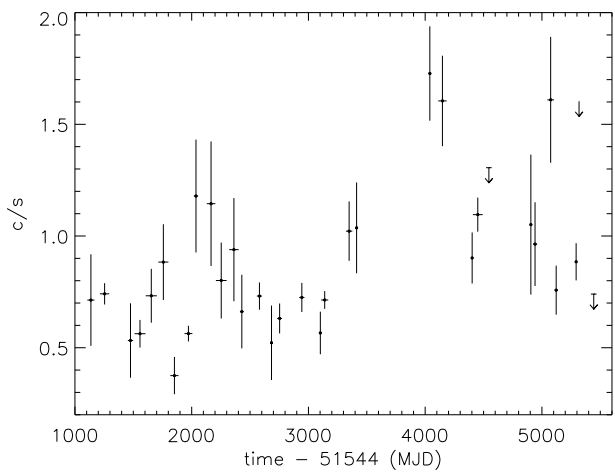}
\end{center}
\caption{22$-$50~keV IBIS/ISGRI lightcurve. 
Error bars correspond to 1$\sigma$ confidence level.
Downward arrows are $3\sigma$ upper-limit.}
\label{figure hr vs flux isgri}
\end{figure}

Figure \ref{figure hr vs flux isgri} shows the IBIS/ISGRI lightcurve of 
\src\ in the energy band 22$-$50~keV.
For each bin (maximum bintime=100\,d) we extracted the flux from the mosaic image.
\src\ is always detected with the exception of three bins, where $3\sigma$ have been reported.
The source shows a significant variability resulting in a high value of $\chi^2=286.27$ (29 d.o.f.)
when the lightcurve is fitted with a constant.

\begin{figure*}
\begin{center}
\includegraphics[bb=80 371 545 695,clip,width=\columnwidth]{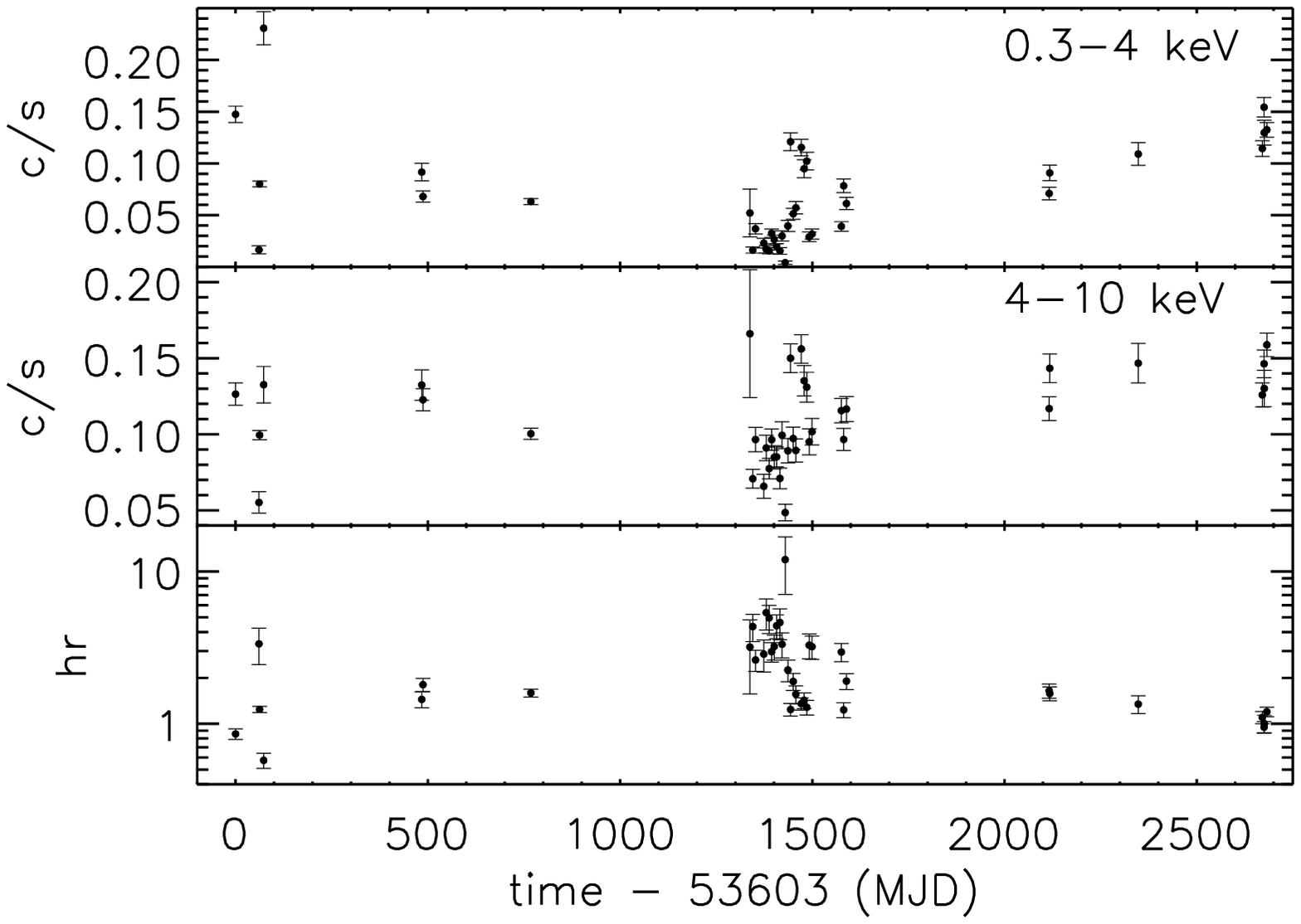}
\includegraphics[bb=80 371 545 695,clip,width=\columnwidth]{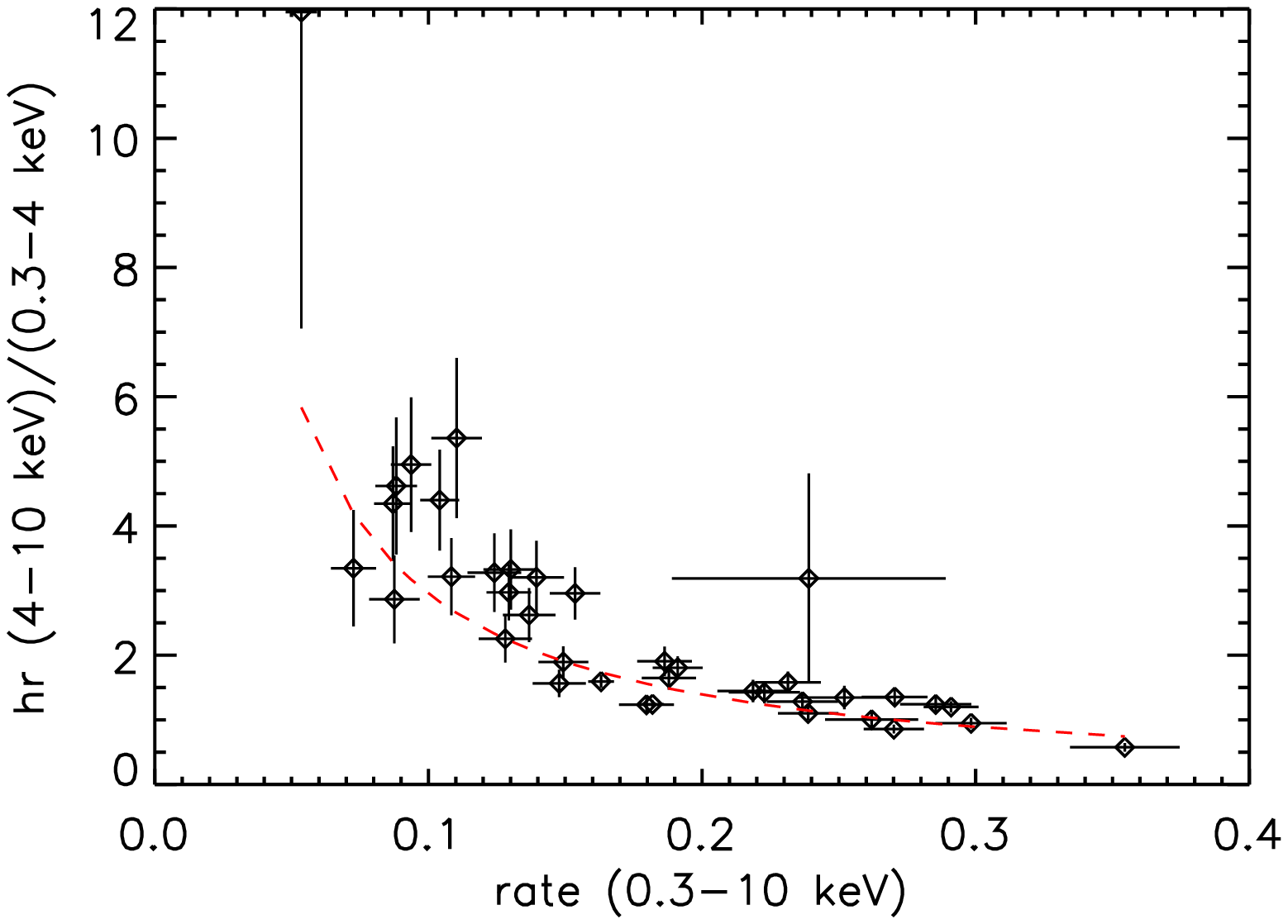}
\end{center}
\caption{\emph{Left panel:} \sw/XRT lightcurves (top: 0.3$-$4~keV; middle: 4$-$10~keV)
and hardness ratios as function of time (bottom panel).
\emph{Right panel} Spectral hardness ratio as function of the 0.3$-$10~keV count rate.
The hardness ratio is the ratio between the 4$-$10~keV count rate
and the 0.3$-$4~keV count rate.
The red dashed line shows the best-fitting power-law model with index $\alpha=-1.09 \pm 0.12$.}
\label{figure hr vs flux}
\end{figure*}

We extracted the \sw/XRT lightcurves of \src\ in two energy bands, 0.3$-$4~keV
and 4$-$10~keV.
They were corrected for point-spread function losses,
vignetting, and background subtracted.
Lightcurves are shown in the left panel of Fig. \ref{figure hr vs flux},
together with their hardness ratio.
Each point corresponds to a single observation.
Right panel of Fig. \ref{figure hr vs flux} shows
the hardness ratio as a function of the 0.3$-$10~keV count rate.
The hardness ratio anti-correlates with the $0.3-10$~keV rate, as previously noted
by \citet{Kennea09} on a sub-sample of \sw/XRT observations.
\citet{Kennea09} showed that the anti-correlation between the flux and hardness ratio
is caused by the variability of the intrinsic absorption.
We used linear, quadratic, and power-law models to fit
the data. The power-law model fits better
the data (index $\alpha = -1.09 \pm 0.12$). 
An F-test comparison of the $\chi^2$ of the linear and power-law
fits gives a probability of chance improvement of the fit of $\sim 1.5\times 10^{-9}$.
The best fit model is plotted in the right panel of Fig. \ref{figure hr vs flux} with a red dashed line.
The fitted parameter $\alpha$ is in good agreement with the value obtained by \citet{Kennea09}.

We searched for coherent pulsations in the $0.3-10$~keV \sw/XRT lightcurve binned at 5~s 
and in the 0.5$-$10~keV \suz\ lightcurve binned at 8~s,
using the fast Lomb-Scargle periodogram technique \citep{Press89}.
The search was performed on time scales between 10~s (\sw) and 16~s (\suz) 
and the time duration of each observation.
The number of independent frequencies was estimated using the equation (13) of \citet{Horne86}.
We found no evidence for pulsations in the data.
We performed simulations on these observations to set a 3$\sigma$ upper-limit
on the pulsed fraction of a sinusoidal signal of about 14\% for \sw/XRT and about 8\% for \suz.

\citet{Luna07} and \citet{Kennea09} searched for coherent periodicity in \chandra\ and \sw/XRT data,
but found no evidence for pulsations.
The pulsed fraction upper-limits derived by \citet{Luna07} and \citet{Kennea09} are in 
agreement with our results.

\subsection{Spectral analysis}
\label{sect. spectral analysis}

We performed the spectral analysis on two set of data:
(\emph{1}) \sw/XRT data jointly with JEM-X and IBIS/ISGRI data,
(\emph{2}) \suz\ data.
We included renormalising constant factors in the spectral fitting
to allow for uncertainties between the instruments
and differences caused by source variability in non simultaneous
observations.
For the cross-normalisation between \suz\ XIS and PIN,
we set the recommended value of 1.18 for observations at the HXD nominal position given in the suzaku-memo-2008-06\footnote{Available at:
\url{http://www.astro.isas.jaxa.jp/suzaku/doc/suzakumemo/suzakumemo-2008-06.pdf}}.

We fitted the spectra with the isobaric cooling flow model ({\tt mkcflow}
in {\tt xspec}\footnote{All spectral fits have been performed
by using {\tt xspec} v12.8.1g \citep{Arnaud96}}), previously used
by \citet{Luna07}, modified by an absorber 
that fully covers the source ({\tt phabs}) and a Gaussian line ({\tt gauss})
for the Fe~K$\alpha$ line at $\sim 6.4$~keV.
This model adequately fits the spectra when two partial covering
absorbers ({\tt pcfabs}) are included. In the framework of the ``clumpy medium''
scenario proposed by \citet{Kennea09}, variable partial covering
components take into account the changes in the absorption
caused by the clumpy absorbing material around the WD of \src\ passing across
the line of sight.

\begin{figure*}
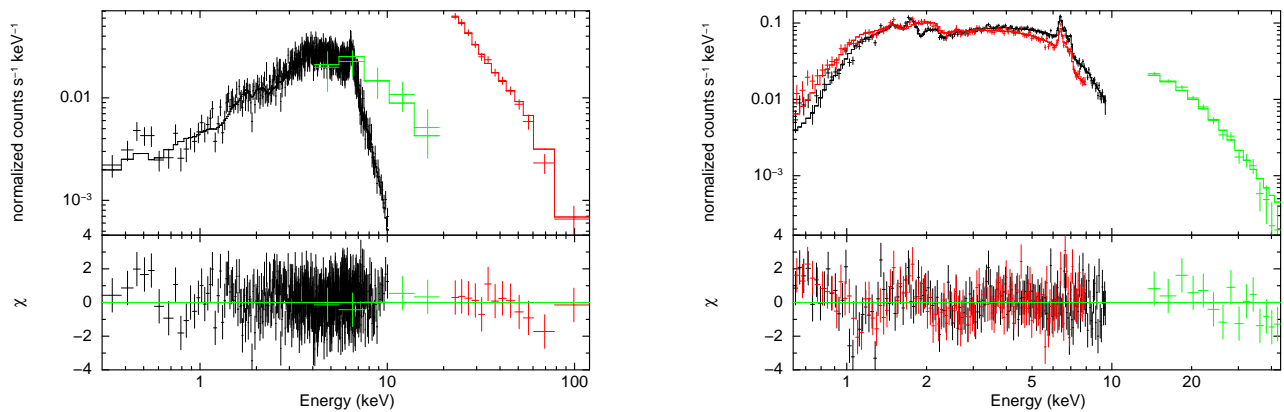

\begin{center}
\includegraphics[angle=-90,width=\columnwidth]{last_mkcflow_xrt_lda.ps}
\includegraphics[angle=-90,width=\columnwidth]{last_mkcflow_suz_lda.ps}
\end{center}
\caption{\emph{Left panel:} \sw/XRT (black), JEM-X1 (green), IBIS/ISGRI (red) spectrum.
\emph{Right panel:} \suz/XIS front-illuminated (black), back-illuminated (red), 
\suz/PIN (green).
For each panel, the residuals in units of standard deviation are shown.
The best fitting parameters are reported in Table \ref{Table spectra}.}
\label{figure spectra}
\end{figure*}

The best-fitting parameters are reported in Table \ref{Table spectra},
the plot of the spectra with residuals in Fig. \ref{figure spectra}.

We find that the absorption and covering fractions vary
between the two data sets, in agreement with the results of \citet{Kennea09}.
Similarly to what found by \citet{Luna07}, we find that the minimum cooling
flow temperature is consistent with the smallest value
allowed by the {\tt mkcflow} model ($kT_{\rm min} = 80.8$\,eV),
while $kT_{\rm max}$ from the spectral fit of the Suzaku data is  pegged at 
the largest value allowed by the model ($kT_{\rm max} = 79.9$\,keV).
The values of the parameter $kT_{\rm max}$ obtained from the spectral fit
of the XRT+INTEGRAL and \suz\ data are different at $90$\% confidence level,
while they are consistent within $99$\% confidence level.
$kT_{\rm max}$ governs the slope of the spectrum at high energies
(the slope increases when $kT_{\rm max}$ decreases) and the overall
height of the spectrum which, however, depends also on the spectral parameter $\dot{M}_{\rm acc}$. 
Therefore, $kT_{\rm max}$ and $\dot{M}_{\rm acc}$ are correlated, 
and a higher $kT_{\rm max}$ in \suz\ spectrum 
(whose coverage at high energies is limited, compared to the INTEGRAL
spectrum, up to $\sim 40$\,keV) 
might not necessarily indicate a significant change of the spectral shape at high energies.
In fact, when we fit the two spectra with a phenomelogical model,
an absorbed power-law modified at energies $E>E_{\rm c}$ with a 
high-energy cutoff ($M(E) = \exp [(E_{\rm c}- E)/E_{\rm f}]$, {\tt highecut} in {\tt xspec}),
we find that the shapes of the spectra are consistent within the uncertainties
(XRT+INTEGRAL: $\Gamma=0.60{+0.10\atop-0.11}$, $E_{\rm c}=5.2{+0.4\atop-0.5}$\,keV, $E_{\rm f}=18\pm2$\,keV;
\suz: $\Gamma=0.77{+0.17\atop-0.12}$, $E_{\rm c}=5.1{+0.8\atop-0.7}$\,keV, $E_{\rm f}=20.1{+4.5\atop-3.3}$\,keV; 
uncertainties are at the 90\% confidence level).
The INTEGRAL spectrum of \src\ has a better coverage at high energies compared to \suz\ data,
hence it is better suited to constrain the parameter $kT_{\rm max}$.

\begin{table}
\begin{center}
\caption{Best-fitting parameters for the \sw/XRT
jointly with the JEM-X and IBIS/ISGRI spectra
and \suz/XIS+PIN (see Fig. \ref{figure spectra}).}
\label{Table spectra}
\begin{tabular}{lcc}
\hline
\hline
\noalign{\smallskip}
Parameters                    & INTEGRAL+XRT               &      Suzaku         \\
\noalign{\smallskip}
\hline
\noalign{\smallskip}
$N_H$ full covering$^a$       & $0.03{+0.02 \atop -0.02}$  &  $0.59{+0.05 \atop -0.05}$   \\
\noalign{\smallskip}
$N_H$ partial covering 1$^a$  & $3.01{+0.09 \atop -0.09}$   &  $4.2{+0.7 \atop -0.6}$  \\
\noalign{\smallskip}
covering fraction 1           & $96.4{+0.5 \atop -0.8}$\%   &  $70{+2 \atop -2}$\%          \\
\noalign{\smallskip}
$N_H$ partial covering 2$^a$  & $13.4{+0.5 \atop -0.5}$     &  $48{+8 \atop -7}$    \\
\noalign{\smallskip}
covering fraction 2           & $81.2{+0.8 \atop -0.8}$\%   & $51{+5 \atop -3}$\%           \\
\noalign{\smallskip}
$kT_{\rm max}$                 & $51{+3 \atop -2}$           & $79.9_{-19.9}$ $^b$      \\
\noalign{\smallskip}
Abundance $^c$                & $0.51{+0.14 \atop -0.15}$  & $0.53{+0.10 \atop -0.09}$     \\
\noalign{\smallskip}
$\dot{M}_{\rm acc}$  $^d$        & $2.00{+0.03 \atop -0.02}$ &  $2.3{-0.8 \atop -0.2}$      \\
\noalign{\smallskip}
$E_{\rm Gauss}$ (keV)          & $6.40 \pm 0.02$        & $6.40\pm 0.01$  \\
\noalign{\smallskip}
$\sigma_{\rm Gauss}$ (keV)     &  $0.13 \pm 0.02$        &  $0.064{+0.019 \atop -0.018}$   \\
\noalign{\smallskip}
norm$_{\rm Gauss}$          & $1.35 \pm 0.15 \times 10^{-4}$ & $8.7 \pm 0.9 \times 10^{5}$ \\ 
\noalign{\smallskip}
EW$_{\rm Gauss}$ (eV)       &  $347{+72 \atop -46}$   &    $211{+68 \atop -22}$           \\
\noalign{\smallskip}
$\chi^2_\nu$ (d.o.f.)    & 1.0448 (523)             &   1.0089 (279)         \\
\noalign{\smallskip}
\hline 
\end{tabular}
\end{center}
Notes. $^a$ Column density in units of $10^{22}$~cm$^{-2}$.
$^b$ Parameter pegged at hard limit; $99$\% lower limit: $51$\,keV.
$^c$ Solar abundance of \citet{Anders89}.
$^d$ Mass accretion rate in units of $10^{-9}$~$M_\odot$~yr$^{-1}$.
Listed uncertainties are at the 90\% confidence level.
\end{table}

\section{Discussion}
\label{sect. discussion}

\citet{Pandel05} studied the X-ray emission from the boundary layers of ten dwarf novae.
They found that the temperature $T_{\rm max}$ of the cooling flow model is related to the
virial temperature $T_{\rm vir}$, and therefore to the mass of the WD 
(e.g. \citealt{Luna07}),
by the equation:
\begin{equation} \label{eq pandel}
  kT_{\rm max} = \frac{3}{5}kT_{\rm vir} = \frac{1}{5}\mu m_{\rm p} v_{\rm K}^2 \mbox{ ,}
\end{equation}
where $v_{\rm K}$ is the keplerian velocity, $\mu$ is the mean molecular weight
and $m_{\rm p}$ is the mass of the proton.
Assuming that the X-ray emission of \src\ is produced in the boundary layer
of an accretion disc around a non-magnetic WD, we used $kT_{\rm max}$
obtained from the spectral fit of the INTEGRAL data described in Sect. \ref{sect. spectral analysis},
the WD mass-radius relation of \citet{Nauenberg72}, and Eq. (\ref{eq pandel})
to obtain an estimate of WD mass of $M_{\rm WD} \approx 1.2$\,M$_\odot$.
This is a refined measurement of the previous estimate obtained by \citet{Luna07} 
($M_{\rm WD} \gtrsim 1.3$\,M$_\odot$) that was based on the $0.3-8$\,keV \emph{Chandra} data
and therefore was highly uncertain due to the limited spectral coverage at high energies.

As mentioned above, \citet{Luna07}
proposed that the X-ray emission of \src\ is produced by 
an optically thin boundary layer of an accretion disc
around a massive non-magnetic WD.
The mass accretion rate obtained in this work (Sect. \ref{sect. spectral analysis}; Table \ref{Table spectra}),
together with the values obtained by \citet{Luna07} and from the formula of the X-ray luminosity of a boundary layer 
$L_{\rm bl} \lesssim 0.5 (GM_{\rm WD}/R_{\rm WD})\dot{M}_{\rm acc}$,
is $\dot{M}_{\rm acc} \approx 2 \times 10^{-9}$~$M_\odot$~yr$^{-1}$.
\citet{Narayan93} found that the boundary layer of a 1~M$_\odot$ WD is expected to be optically thin
for $\dot{M}_{\rm acc} \lesssim 3 \times 10^{-10}$~M$_\odot$~yr$^{-1}$.
Therefore, for high mass accretion rates as that of \src, 
the boundary layer is expected to be optically thick
and the X-ray spectrum soft, with $kT \lesssim 0.1$~keV 
(\citealt{Pringle77}; \citealt{FKR})\footnote{Some cataclysmic variables with an 
optically thick boundary layer show, in addition to the bright soft component, a residual hard X-ray emission
whose luminosity and temperature are lower in comparison to those observed
when their boundary layers are optically thin (e.g. \citealt{Wheatley03}).}.
The transition from optically thick to optically thin boundary layer solutions 
depends on $\dot{M}_{\rm acc}$, the mass and the rotational angular velocity of the WD, 
and the viscosity parameter $\alpha$ of the accretion disc \citep{Popham95}.
Although its exact location in the space of parameters
is still not well known, \citet{Popham95} found that it occurs at larger
values of $\dot{M}_{\rm acc}$ for a more massive WD.
On the other hand, \citet{Suleimanov14} noted that
the transition from optically thick to optically thin boundary layer
should occur at lower values of $\dot{M}_{\rm acc}$ (compared to the results of \citealt{Popham95})
if a more realistic coefficient for the Kramers opacity is used.
Although the optically thin boundary layer scenario cannot be excluded,
its presence in \src\ is questionable. Therefore,
it is worth considering alternative explanations
for the hard X-ray emission of \src.

An alternative scenario 
to explain the observed hard X-ray emission is
the accretion onto the polar caps of a magnetised WD. 
In these objects, the magnetic field is strong enough to disrupt the accretion flow geometry
(which can be quasi-spherical or an accretion disc) at some distance from the star surface,
and channel the matter onto the polar caps.
Shocks form above the polar caps and the hot, optically-thin post-shock region (PSR) emits
hard X-ray photons.
Assuming that the compact object of \src\ is a magnetic WD,
we fitted the XRT+INTEGRAL and \suz\ spectra with the PSR model of \citet{Suleimanov05}
modified by an absorber that fully cover the source, one partial covering absorber,
and three narrow Gaussians for the iron-line complex.
The PSR model of \citet{Suleimanov05}, developed for intermediate polars, 
allows to estimate the mass of the WD from the temperature 
and emissivity distribution in the post-shock region.
We fitted the spectra above $3$\,keV, i.e. in the energy range of validity of the model of \citet{Suleimanov05}.
The best fits of the spectra give 
$M_{\rm WD} = 0.93{+0.08 \atop -0.07}$~M$_\odot$ for XRT+INTEGRAL data set
and $M_{\rm WD} =1.13{+0.15 \atop -0.15}$~M$_\odot$ for \suz\ spectrum 
(uncertainties are at the 90\% confidence level).

The magnetic WD scenario explains reasonably well the 
main properties of the X-ray emission of \src.
In fact, accreting WDs with strong magnetic fields at the surface ($B\sim 10^6-10^7$~G)
are observed at high X-ray luminosities ($10^{32}-10^{34}$~erg~s$^{-1}$; e.g. \citealt{Sazonov06})
and with hard X-ray spectra, that can be observed with the currently operating
hard X-ray telescopes up to $\sim 100$~keV (e.g. \citealt{Scaringi10} and references therein).

The X-ray emission of magnetic WDs is pulsed with typical pulsed fractions in the 40\%$-$90\% range
(\citealt{Kennea09} and references therein).
If \src\ is a magnetic WD, the lack of detection of the pulsation can be explained 
with a relatively low pulsed fraction of the signal, caused,
for example, by aligned or nearly aligned rotation and magnetic axes. 
The aligned axes argument was proposed by \citet{Ramsay08}
to explain the lack of detection of coherent X-ray pulsations in V426\,Oph and LS\,Peg,
two cataclysmic variables with properties which are similar to intermediate polars.
They also suggest that some of the unclassified cataclysmic variables
with hard X-ray emission ($20-100$\,keV) that do not show a spin period modulation
might be intermediate polars with closely aligned magnetic and spin axes.

\section{Conclusions}

We presented a study of the spectral X-ray emission of \src\
based on $\sim8.6$~Ms of archival \inte\ data,
$\sim140$~ks of \sw/XRT data, and a \suz\ observation of 39\,k.
We found that a cooling flow model ({\tt mkcflow})
modified by an absorber that fully covers the source,
two partial covering absorbers and a Gaussian line
for the Fe~K$\alpha$ line at $\sim 6.4$~keV
adequately fits the X-ray spectrum.
We found that the absorption components are variables, 
in agreement with the ``clumpy medium'' scenario proposed by \citet{Kennea09}.
So far, the origin of the hard X-ray emission of \src\ has been explained with 
an optically thin boundary layer around a non-magnetic WD.
In this framework, we used the relation between the $kT_{\rm max}$ temperature
and the mass of the WD of \citet{Pandel05} to refine the previous measurement 
of the mass of the WD obtained by \citet{Luna07} to the new value of $M_{\rm WD} \approx 1.2$~M$_\odot$.
Since the mass accretion rate might be too high for the optically thin boundary layer scenario,
we have shown that another plausible scenario is that the compact object 
is a magnetic WD, with the hard X-ray emission produced 
from the post-shock region above the polar caps.
Using the spectral model of \citet{Suleimanov05}, 
we estimated a mass of the WD of $\sim 0.9-1.1$~M$_\odot$.
Furthermore, we used 2MASS photometry to estimate the distance to the source of $1.2-2.4$\,kpc.

\begin{acknowledgements}
We thank the anonymous referee for constructive comments, 
which helped to improve the paper considerably.
This work is supported by the Bundesministerium f\"ur
Wirtschaft und Technologie through the Deutsches Zentrum f\"ur Luft
und Raumfahrt (grant FKZ 50 OG 1602).
V.D. thanks the Deutsches Zentrums f\"ur Luft
und Raumfahrt (DLR) for financial support (grant DLR 50 OR 0702).
V.S. thanks DFG for financial support (grant WE 1312/48-1).
This work is partially supported by the Polish NCN grant 2012/04/M/ST9/00780.
This paper is based on data from observations with INTEGRAL, 
an ESA project with instruments and science data centre funded by ESA
member states (especially the PI countries: Denmark, France, Germany,
Italy, Spain, and Switzerland), Czech Republic and Poland,
and with the participation of Russia and the USA.
This research has made use of data obtained from the Suzaku satellite, 
a collaborative mission between the space agencies of Japan (JAXA) 
and the USA (NASA). 

\end{acknowledgements}

\bibliographystyle{aa} 
\bibliography{lducci_rtcru}

\end{document}